\documentclass[prl,showpacs,preprintnumbers,amsmath,amssymb,tightenlines,twocolumn,superscriptaddress]{revtex4}

\usepackage{graphicx}
\usepackage{dcolumn}
\usepackage{bm}
\usepackage{epsfig}

\newcommand{\ssection}[1]{{\noi  \it #1:}}
\newcommand{\bra}[1]{\langle\,{#1}\, |}
\newcommand{\ket}[1]{|\,{#1}\,\rangle}

\newcommand{\expec}[1]{\langle #1 \rangle}

\newcommand{\sub}[2]{{#1}_{\mbox{\!\! \scriptsize #2}}}
\newcommand{\supp}[2]{{#1}^{\mbox{\!\! \scriptsize #2}}}

\newcommand{\bv}[1]{\mathbf{ #1 }}

\def\noi{\noindent}
\def\beq{\begin{equation}}
\def\eeq{\end{equation}}

\def\CR{\nonumber\\[0.15cm]}

\def\figurewidth{0.99}
\newcommand{\rref}[1]{Ref.~\cite{#1}}
\newcommand{\fref}[1]{Fig.~\ref{#1}}
\newcommand{\frefp}[2]{Fig.~\ref{#1}~(#2)}

\newcommand{\eref}[1]{Eq.~(\ref{#1})}

\newcommand{\essref}[2]{Eqs.~(\ref{#1})-(\ref{#2})}

\newcommand{\cref}[1]{chapter~\ref{#1}}

\newcommand{\Cref}[1]{Chapter~\ref{#1}}

\newcommand{\bref}[1]{(\ref{#1})}

\usepackage{ulem}  
\usepackage[usenames,dvipsnames]{color}

\begin{document}

\title{Quantum simulation of energy transport with embedded Rydberg aggregates}

\author{D.~W.~Sch{\"o}nleber}
\affiliation{Max Planck Institute for the Physics of Complex Systems, N\"othnitzer Strasse 38, 01187 Dresden, Germany}
\author{A.~Eisfeld}
\affiliation{Max Planck Institute for the Physics of Complex Systems, N\"othnitzer Strasse 38, 01187 Dresden, Germany}
\author{M.~Genkin}
\affiliation{Max Planck Institute for the Physics of Complex Systems, N\"othnitzer Strasse 38, 01187 Dresden, Germany}
\author{S.~Whitlock}
\affiliation{Physikalisches Institut, Universit{\"a}t Heidelberg, Im Neuenheimer Feld 226, 69120, Heidelberg, Germany}
\author{S.~W\"uster}
\affiliation{Max Planck Institute for the Physics of Complex Systems, N\"othnitzer Strasse 38, 01187 Dresden, Germany}
\email{sew654@pks.mpg.de}
\begin{abstract}
We show that an array of ultracold Rydberg atoms embedded in a laser driven background gas can serve as an aggregate for simulating 
exciton dynamics and energy transport with a controlled environment. Spatial disorder and decoherence introduced by the interaction with the background gas 
atoms can be controlled by the laser parameters. This allows for an almost ideal realization of a Haken-Reineker-Strobl type model for energy transport. Physics can be monitored using the same mechanism that provides control over the environment.
The degree of decoherence is traced back to information gained on the excitation location through the monitoring, turning the setup into an experimentally accessible model system for studying the effects of quantum measurements on the dynamics of a many-body quantum system. 
\end{abstract}

\maketitle

\ssection{Introduction}
%
Excitation transport through dipole-dipole interactions \cite{FrTe38_861_,frenkel_exciton} plays a prominent role in diverse physical settings, including 
photosynthesis \cite{grondelle:review,grondelle:book}, exciton transport through quantum-dot arrays \cite{kagan:longrange:QD}, and molecular aggregates 
\cite{saikin:excitonreview,kirstein:Jaggregates,kuehn_lochbrunner:excitonquantdyn}.  Of crucial importance is the competition between the fundamentally coherent transport mechanism and the 
coupling to the environment, which has been under intense scrutiny in the context of photosynthesis 
(e.g.~\cite{renger:review,Foe48_55_,Foe65_93_,FrTe38_861_,Sc37_795_,AmVaGr00__}) and recently experienced a resurge of interest 
(e.g.~\cite{engel_fleming:coherence_nature,PlHu08_113019_,
MoReLl08_174106_,MiBrGr10_257_,
scholak:optimizedFMO,scholak:randomFMO}). Often, clean studies of excitation transport are impeded by the large number of degrees of freedom in these systems, 
for example, strongly coupled vibrational modes \cite{renger:review,RiRoSt11_113034_}. Ultracold atoms  prepared in highly-excited Rydberg states exhibit 
similar dipolar state-changing interactions~\cite{anderson:resonant_dipole,noordam:interactions,book:gallagher,park:dipdipionization,barredo:trimeragg,ravets:foersterdipdip} as found 
in organic molecules, but are considerably simpler to study. Due to their strong interactions and relative ease to control using lasers, Rydberg atoms have been 
proposed as quantum simulators for quantum spin models \cite{weimer:quantsim,Lesanovsky:correlspinchain} and electron-phonon interactions 
\cite{hague:Su_Schrieffer_Heeger}. Aggregates formed by networks of Rydberg atoms (Rydberg aggregates) 
\cite{eisfeld:Jagg,muelken:excitontransfer} are also ideally suited to the study of dipolar energy transport in an experimentally accessible system, as recently 
demonstrated~\cite{guenter:EITexpt}. 

Here we study energy transport through a Rydberg aggregate embedded within an optically driven background gas that acts as a precisely controlled environment. 
This system extends the one recently used to observe diffusive excitation  transport~\cite{guenter:EITexpt} by separating the aggregate degrees of freedom from 
those of the background gas. The background gas is electromagnetically rendered transparent for a probe beam. Only in the vicinity of the aggregate atoms, interactions disrupt this 
transparency, causing each aggregate atom to cast a shadow with radius given by the interaction strength. We demonstrate parameters for which a larger 
absorption shadow is cast by the atom carrying an excitation, allowing us to infer its location.
\begin{figure}[t]
\centering
\epsfig{file=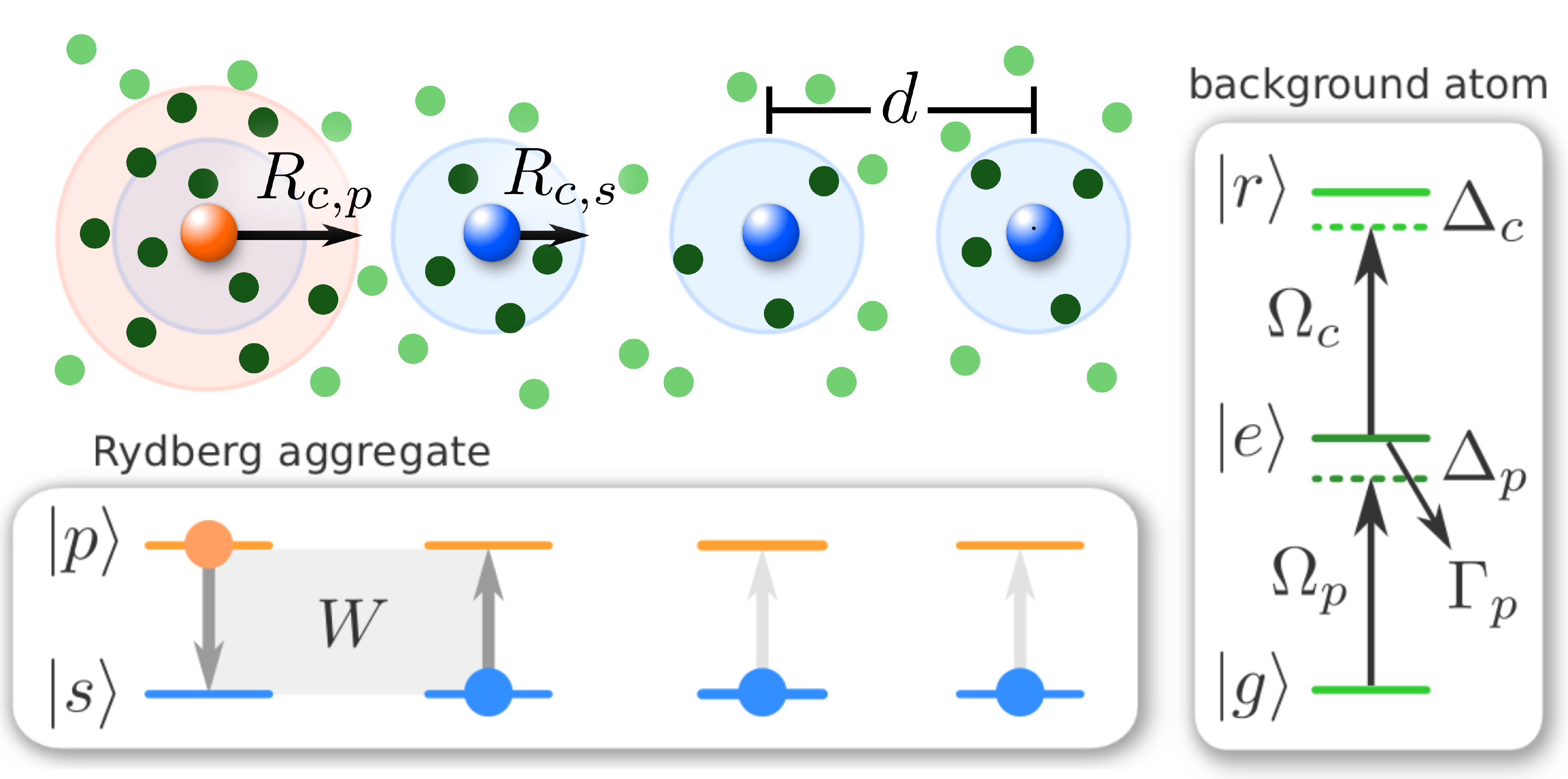,width=\figurewidth\columnwidth} 
\caption{(color online) Sketch of an embedded Rydberg aggregate. An assembly of several Rydberg atoms in a state $\ket{s}$ (large blue) 
and one in a state $\ket{p}$ (large orange) is linearly arranged with spacing $d$ in a background atomic gas (shades of green). These background atoms are then 
addressed with an EIT scheme (right panel), providing detection signals within radii $R_{c,s/p}$ around each aggregate atom.  
\label{embedding_setup}}
\end{figure}
We show that the background gas simultaneously causes a back-action on the aggregate which can give rise to non-Gaussian disorder as well as 
site-dependent dephasing. The resulting excitation transfer dynamics can be described by a master equation similar to the one introduced by Haken-Reineker-Strobl (HRS) \cite{HaRe71_253_,HaSt73_135_,EiBr12_046118_} to study the transition from coherent to incoherent transport.

The experimental realization of a {controllable} HRS-type model will {benefit the study of excitation transport in an open system, be it semi-conductors or  light harvesting}. For the latter extensions to exciton-vibrational coupling 
and non-Markovian environments may be required \cite{spano:exciton_vibr_coupl,renger:review,eisfeld:Jagg,RiRoSt11_113034_,mostame:supercond}. Finally, we show 
how decoherence in this system is intimately linked to the information obtained by the background gas acting as a quantum measurement device. In particular, 
despite strong aggregate-background interactions, decoherence vanishes if the background 
atoms do not allow one to infer the location of the excitation.

\ssection{Scheme and model}
%
The system we propose consists of a chain of $N$ Rydberg atoms with spacing $d$ forming the aggregate sketched in \fref{embedding_setup}. Such an 
arrangement can be created by exciting Rydberg states from a trapped ultracold atomic gas using tightly focused laser beams 
\cite{nogrette:hologarrays,ScTiKr11_907_}, or by pulsed or chirped excitation in the dipole blockade regime \cite{jaksch:dipoleblockade,lukin:quantuminfo}, 
which gives rise to spatially correlated Rydberg excitation patterns~\cite{pohl:crystal,gaerttner:floatcrystals,
weimer:phases,wuester:echo,lesanovsky:nonequil_structures,cenap:emergent,rvb:crystalchirped,schempp:rydaggstatistics,schauss:mesocryst,schauss:dyncryst}.  
$N-1$ atoms are initially prepared in the state $\ket{s}=\ket{\nu s}$ with principal quantum number $\nu$ and angular momentum $l=0$, while a single atom is 
excited to the state $\ket{p}=\ket{\nu p}$, with angular momentum $l=1$.  This $\ket{p}$ excitation can then migrate through the aggregate through resonant 
dipole-dipole exchange interactions~\cite{book:gallagher,noordam:interactions}. In addition, the aggregate is immersed in a gas of $M$ background atoms, initially prepared 
in the electronic ground state $\ket{g}$,  the positions of which could be random or arranged in a regular fashion. These atoms are coupled by two laser fields from $\ket{g}$ 
via a short-lived intermediate state $\ket{e}$ 
(spontaneous decay rate $\Gamma_p$) to a third Rydberg level, $\ket{r}=\ket{\nu' 
s}$~\cite{friedler:longrangepulses,Mohapatra:coherent_ryd_det,mauger:strontspec,Mohapatra:giantelectroopt,schempp:cpt,sevincli:quantuminterf,
parigi:interactionnonlin}. Aggregate and background atoms could be the same or different atomic species.

This system is governed by the many-body Lindblad master equation for the density matrix $\hat{\rho}$ ($h=1$)
\begin{align}
\dot{\hat{\rho}}= -i [\hat{H},\hat{\rho}] + \sum_\alpha {\cal L}_{\hat{L}_\alpha}[\hat{\rho}].
\label{mastereqn}
\end{align}
The Hamiltonian consists of three parts, $\hat{H}=\sub{\hat{H}}{agg}+\sub{\hat{H}}{EIT}+\sub{\hat{H}}{int}$, for the aggregate, the 
background gas of three-level atoms and van-der-Waals (vdW) interactions \cite{book:gallagher,singer:VdWcoefficients,beguin:vdWmeasure} between atoms that are in a Rydberg state. The super-operator ${\cal 
L}_{\hat{L}_\alpha}[\hat{\rho}]$ describes spontaneous 
decay of the background atom $\alpha$ from level $\ket{e}$, thus 
 $ {\cal L}_{\hat{O}}[\hat{\rho}]=\hat{O}\hat{\rho}\hat{O}^\dagger - (\hat{O}^\dagger\hat{O}\hat{\rho}+\hat{\rho}\hat{O}^\dagger\hat{O})/2$ and the decay 
operator is $\hat{L}_\alpha=\sqrt{\Gamma_p}\hat{\sigma}^{(\alpha)}_{ge}$, with $\hat{\sigma}^{(\alpha)}_{kk'}=[\ket{k}\bra{k'}]_\alpha$ acting on atom $\alpha$ 
only and $k,k'\in \{g,e,r,s,p\}$.
  
The aggregate atoms are labeled by Latin indices such as $n$ and $m$. Restricted to the Hilbert space with a single excitation and setting the constant 
energy splitting between $s$ and $p$ to zero, we can write
\begin{align}
\sub{\hat{H}}{agg}&= \sum_{n\neq m} W_{nm} \hat{\sigma}^{(n)}_{sp}\hat{\sigma}^{(m)}_{ps}= \sum_{n\neq m} W_{nm} \ket{\pi_n}\bra{\pi_m},
\label{Hagg}
\end{align}
where $\ket{\pi_n}=|ss..p..ss\rangle$ (all aggregate atoms are in $\ket{s}$ except the $n$'th, which is in $\ket{p}$) and $W_{nm}=C_3/|r_n - r_m|^3$. Here 
$C_3$ is the dipole-dipole interaction strength and $r_{n}$ is the position of aggregate atom $n$. We call eigenstates of \bref{Hagg} excitons 
\cite{book:maykuehn,cenap:motion}. For simplicity we have ignored vdW interactions between aggregate atoms~\cite{hashem:VdW}.

The Hamiltonian for the background gas in the rotating wave approximation reads
\begin{align}
\sub{\hat{H}}{EIT}= \sum_\alpha  &\bigg[\frac{\sub{\Omega}{p}}{2}\hat{\sigma}^{(\alpha)}_{eg}  + \frac{\sub{\Omega}{c}}{2}\hat{\sigma}^{(\alpha)}_{re}  + \mbox{h.c.} 
\CR
&-\Delta_p\hat{\sigma}^{(\alpha)}_{ee}-(\Delta_p + \Delta_c)\hat{\sigma}^{(\alpha)}_{rr}     \bigg],
\label{Heit}
\end{align}
where $\Omega_{p,c}$ and $\Delta_{p,c}$ are the probe and coupling Rabi frequencies and detunings respectively. Typically $\Omega_p \ll \Omega_c$ and $\Delta_p + \Delta_c=0$ which corresponds to conditions of electromagnetically induced transparency (EIT) used for Rydberg atom detection \cite{guenter:EIT,guenter:EITexpt}.

Background atoms interact among themselves and with the aggregate through vdW interactions
\begin{align}
\sub{\hat{H}}{int}&= \sum_{\alpha<\beta} V_{\alpha\beta}^{(rr)} \hat{\sigma}^{(\alpha)}_{rr}\hat{\sigma}^{(\beta)}_{rr} +  \sum_{a\in\{s,p\},\alpha n } \!\! V_{\alpha n}^{(ra)} \hat{\sigma}^{(\alpha)}_{rr}\hat{\sigma}^{(n)}_{aa}.
\label{Hint}
\end{align}
For simplicity we assume isotropic interactions. To use the background gas as a probe for the state of the aggregate it is necessary that the interactions are 
state dependent. The interaction strength between two atoms $\alpha$, $n$ is $V_{{\alpha}n}^{(ra)}=V^{(ra)}(|r_\alpha - r_n|)=C_{\eta(a),ra}/|r_\alpha - 
r_n|^{\eta(a)}$, when they are in states $\ket{r}$ and $\ket{a}\in\{\ket{s},\ket{p} \}$. As concrete examples we consider $\ket{s}=\ket{43s}$, 
$\ket{p}=\ket{43p}$ in $^{87}$Rb, with two choices $\ket{r}=\ket{38s}$ or $\ket{r'}=\ket{17s}$ for the upper state of the EIT ladder. The former realizes 
power laws $\eta(a)=6,6,4$ for $a=r,s,p$, respectively, with $|V^{(rp)}| \gg |V^{(rs)}|$, due to a nearly resonant process: $43 p+38 s \leftrightarrow 41 d+38p$ 
\cite{footnote:interactions}, and the latter has $\eta(p)=6$ and $V^{(rp)} \approx V^{(rs)}$. 

\ssection{Excitation detection}
On resonance ($\Delta_{p,c}=0$), the background gas becomes transparent for the probe beam described by $\Omega_p$. However, close to the aggregate atoms, 
interactions $V_{\alpha n}^{(ra)}> V_c=\Omega_c^2/(2\Gamma_p)$ destroy the transparency \cite{guenter:EIT,guenter:EITexpt} (see also 
\cite{olmos:amplification}).
This creates an absorption shadow around each aggregate atom, the radius $R_{c,a}=(2 C_{\eta(a),ra} \Gamma_p/\Omega_c^2)^{1/\eta(a)}$ of which 
depends on the state $a\in\{s,p\}$, as sketched by blue (orange) circles in \fref{embedding_setup}. Through this difference we can infer the
location of the $p$-excitation.

\ssection{Effective aggregate model}
To derive an effective model for the aggregate alone we proceed by adiabatically eliminating the internal states of the background atoms following the approach described in 
Ref.~\cite{sorensen:adiabelim}. This is justified when 
the time scale on which background atoms would approach a steady state, set by the atomic decay rate $1/\Gamma_p$,
is shorter than that for excitation transport $1/W(d)=d^3/C_3$ (for details see \cite{sup:info}). The evolution of the reduced aggregate density matrix 
$\supp{\hat{\rho}}{(agg)}=\sum_{nm}\rho_{nm}\ket{\pi_n}\bra{\pi_m}$, in the case $\Delta_{p,c}=0$, $V^{(rr)}_{\alpha\beta}=0$ and to leading order in $\Omega_p$, obeys:
\begin{align}
\supp{\dot{\hat{\rho}}}{(agg)}&=-i[\sub{\hat{H}}{agg} + \sub{\hat{H}}{eff},\supp{\hat{\rho}}{(agg)}]+
 \sum_\alpha {\cal L}_{\sub{\hat{L}}{eff}^{(\alpha)}  } [\supp{\hat{\rho}}{(agg)}],
\label{effective_eqn}
\\
\sub{\hat{H}}{eff}&=  \sum_{n}\left[ \sum_\alpha \frac{ \Omega_p^2}{ \Omega_c^2 } \frac{    \bar{V}_{n\alpha} }{1 +    (\bar{V}_{n\alpha}/\sub{V}{c})^2   }   \right]\ket{\pi_n}\bra{\pi_n},
\label{effective_hamil}
\\
\sub{\hat{L}}{eff}^{(\alpha)}&= \sum_n \left[  \frac{ \Omega_p}{ \sqrt{\Gamma_p} } \frac{1 }{i +    \sub{V}{c}/\bar{V}_{n\alpha}  } \right]   
\ket{\pi_n}\bra{\pi_n},
\label{effective_dephasing}
\end{align}
where we have introduced the background-aggregate interaction $\bar{V}_{n\alpha} = V^{(rp)}_{n\alpha}+ \sum_{m\neq n} V^{(rs)}_{m\alpha}$. Note the imaginary contributions to $\sub{\hat{L}}{eff}^{(\alpha)}$. For the case $\Delta_{p,c}\neq0$, see \cite{sup:info}.

The effective Hamiltonian \bref{effective_hamil} describes a mean energy shift of aggregate site $n$ due to the interaction with the level $\ket{r}$ of the
background atoms, weighted by the steady-state occupation of $\ket{r}$. The strength of the second term \bref{effective_dephasing} is set by the two-level atom photon scattering rate
$\sub{\gamma}{eff}\approx\Omega_p^2 /\Gamma_p$ within the critical radius of an aggregate atom. Imaginary off-diagonal terms in \bref{effective_eqn} arising 
from  imaginary parts of \bref{effective_dephasing} can be interpreted as a contribution to the disorder \cite{sup:info}, while real ones describe dephasing mechanisms. The relative 
contributions of disorder and dephasing terms can be controlled by choosing Rydberg states with different interactions and through the EIT laser parameters. 
\eref{effective_eqn} furnishes a Haken-Reineker-Strobl type model \cite{EiBr12_046118_} for excitation 
transport. All scenarios from dominant dephasing to dominant disorder can be realized by varying the intermediate state detuning $\Delta_p$ while keeping the 
two-photon detuning fixed: $\Delta_p+\Delta_c\approx0$. In particular, for large $\Delta_p$ the contribution of dephasing can be significantly reduced, see \cite{sup:info}.

\begin{figure}[t!]
\centering
\epsfig{file=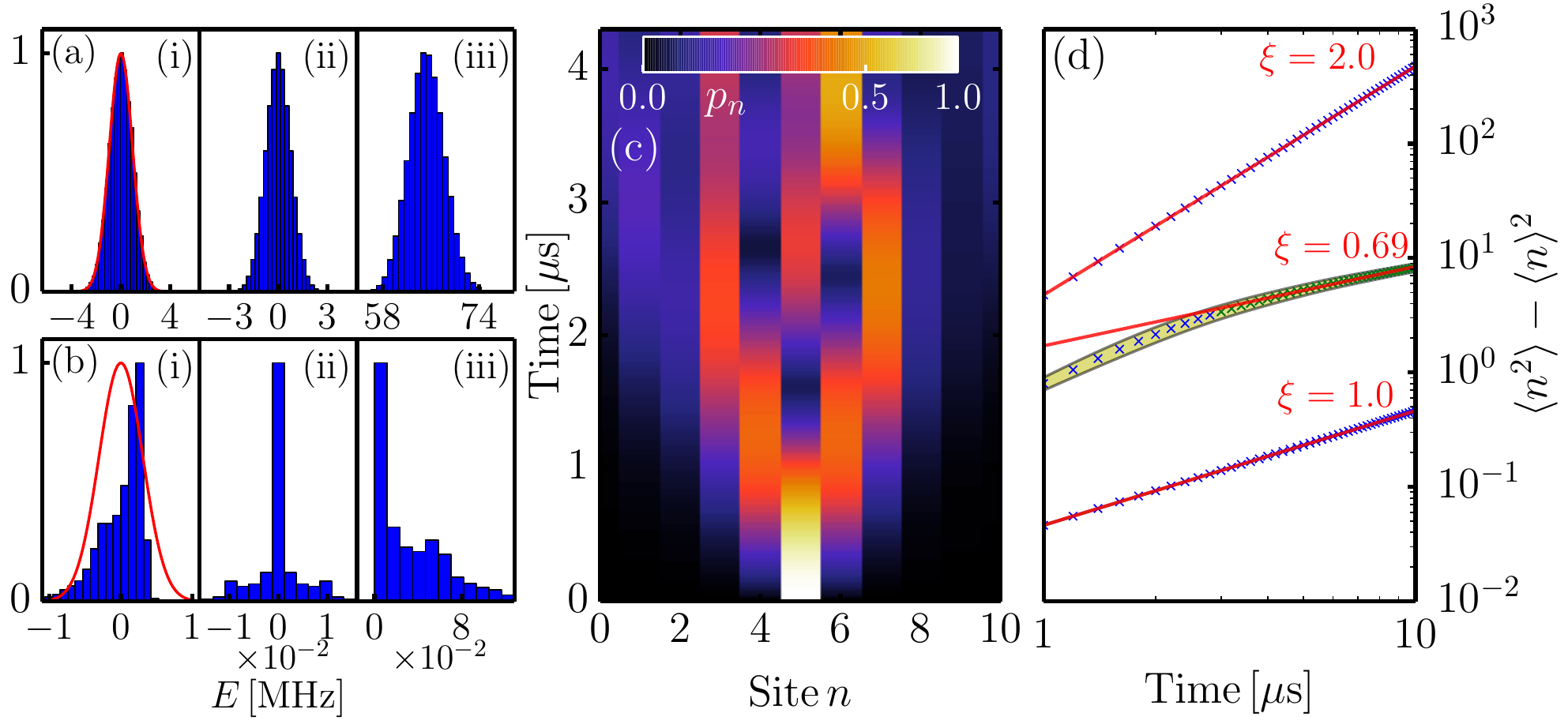,width= \figurewidth\columnwidth} 
\caption{(color online) Varying disorder and dephasing in quantum simulations of energy transport. 
(a) Histograms for disorder and dephasing (i) $P_{E}$, (ii) $P_{\epsilon}$, (iii) $P_\gamma$, using parameters $\Omega_p=1.3$ MHz, $\Omega_c=30$ MHz, 
$\Delta_p=\Delta_c=0$,  $\Gamma_p=6.1$ MHz, $d=19{\mu}$m, $W(d)=0.24$ MHz, $\sub{\rho}{bg}=3.8\times10^{18}$ m$^{-3}$, $C_{4,rp} = -1032$ MHz ${\mu}$m$^4$, 
$C_{6,rs} = -87$ MHz $\mu$m$^6$, thus assuming an upper background level $\ket{r}$.
(b) The same as (a), with parameters $\Omega_p=12$ MHz, $\Omega_c=90$ MHz, $\Delta_p=-20$ MHz, $\Delta_c=22$ MHz,  $d=24{\mu}$m, $W(d)=0.1$ MHz, 
$\sub{\rho}{bg}=9.5\times10^{17}$ m$^{-3}$, $C_{6,rp} = -0.4$ MHz ${\mu}$m$^6$, $C_{6,rs} = -0.1$ MHz $\mu$m$^6$, thus assuming an upper background level 
$\ket{r'}$. 
(c) Effect of disorder on a single transport realization, using parameters as in (b). (d) Variance of the excitation location ($\times$), fit by 
$\sigma_n^2(t)= S t^\xi$. From bottom to top parameters as in (a), (b), (a) with $\Omega_p=0$.
\label{long_chain}}
\end{figure}

In the following we analyze the influence of the disorder and dephasing introduced by the background gas. More explicitly \eref{effective_eqn} reads
$\dot{\rho}_{nm}= \sum_{k} i (W_{km} \rho_{nk} - W_{nk} \rho_{km}) 
+ i (E_m-E_n+ \epsilon_{nm}) \rho_{nm} - \gamma_{nm}\,\rho_{nm}/2$,
with
$E_n = \sum_\alpha \sub{H}{eff}^{(n\alpha)}$, $\epsilon_{nm}= \sum_\alpha\mbox{\cal Im}[\sub{L}{eff}^{(n\alpha)}\sub{L}{eff}^{(m\alpha)*}]$, and
$ \gamma_{nm}= \sum_\alpha(|\sub{L}{eff}^{(n\alpha)}|^2 + |\sub{L}{eff}^{(m\alpha)}|^2 -2\mbox{\cal Re}[\sub{L}{eff}^{(n\alpha)}\sub{L}{eff}^{(m\alpha)*}] )$.
We then define distributions  $P_E(E_n - \langle E_n \rangle)$, $P_\epsilon(\epsilon_{nm} - \langle \epsilon_{nm} \rangle)$ and $P_\gamma(\gamma_{nm})$, for the probability with which an individual background atom $\alpha$ contributes to 
disorder and dephasing in an ensemble average over background atom positions ($E_n$ disorder from \bref{effective_hamil}, $\epsilon_{nm}$ disorder 
from \bref{effective_dephasing}, $P_\gamma$ dephasing). Both the width and 
the shape of these distributions can be controlled by the  laser parameters and the background atom density. In \fref{long_chain} we show two examples: panel (a) 
corresponds to resonant EIT excitation ($\Delta_{p,c}=0$) and large background gas density, resulting in dominant dephasing and Gaussian 
distributions, while panel (b) shows the case of finite intermediate state detuning and low density such that interactions between probe and aggregate atoms are 
weaker. Detuning from the intermediate level reduces spontaneous decay and makes dephasing weaker than disorder. Remarkably, for low densities and weak interactions we find
significant outliers in the atomic distance distribution that cause non-Gaussian disorder which can crucially modify excitation transport 
\cite{bas:levy}. By controlling the placement of individual background atoms using microstructured optical traps, even more exotic forms of disorder could be studied. 

The effects of disorder and dephasing on transport can be seen in \frefp{long_chain}{c}, where we show a single realization of \eref{effective_eqn} for $N=11$ atoms immersed 
in a gas of randomly but homogeneously distributed background atoms. In a corresponding ensemble average, the spatial width of the excitation distribution over 
aggregate sites $\sigma_n^2 =\expec{n^2}-\expec{n}^2$ carries the transport signatures \frefp{long_chain}{d}. Parametrizing $\sigma_n^2(t)= S t^\xi$, we find 
$\xi=2$ for ballistic transport ($\Omega_p=0$), $\xi=1$ for typical diffusive transport resulting from \frefp{long_chain}{a} and $\xi=0.69$ for 
sub-diffusive transport arising from the non-Gaussian disorder in \frefp{long_chain}{b}. 

\ssection{Imaging and measurement-induced decoherence}
%
The degree of decoherence present in this system is intimately linked to the action of the background gas acting as a real-time probe of the aggregate, 
making it an appealing model to demonstrate measurement-induced decoherence \cite{korotkov:weak_measurement_deph}. Since the background gas degrees of freedom have been eliminated in the effective model, we demonstrate this effect with simulations of the full master equation, which also serve to verify model \bref{effective_eqn}. We study an aggregate with $N=3$, probed by two randomly distributed pairs of background atoms, using a quantum-jump Monte-Carlo technique \cite{molmer:quantumjump,david:cohincoh}. The background atom pairs have 
a separation ${\Delta}r=0.3$ $\mu$m, yielding $V^{(rr)}({\Delta}r)=730$ GHz \cite{footnote:cutoff} to include significant interactions between background atoms. 
We initially prepare the aggregate in state $\ket{\pi_1}$ and all background atoms in their ground state $\ket{g}$. 

\begin{figure}[t!]
\centering
\epsfig{file=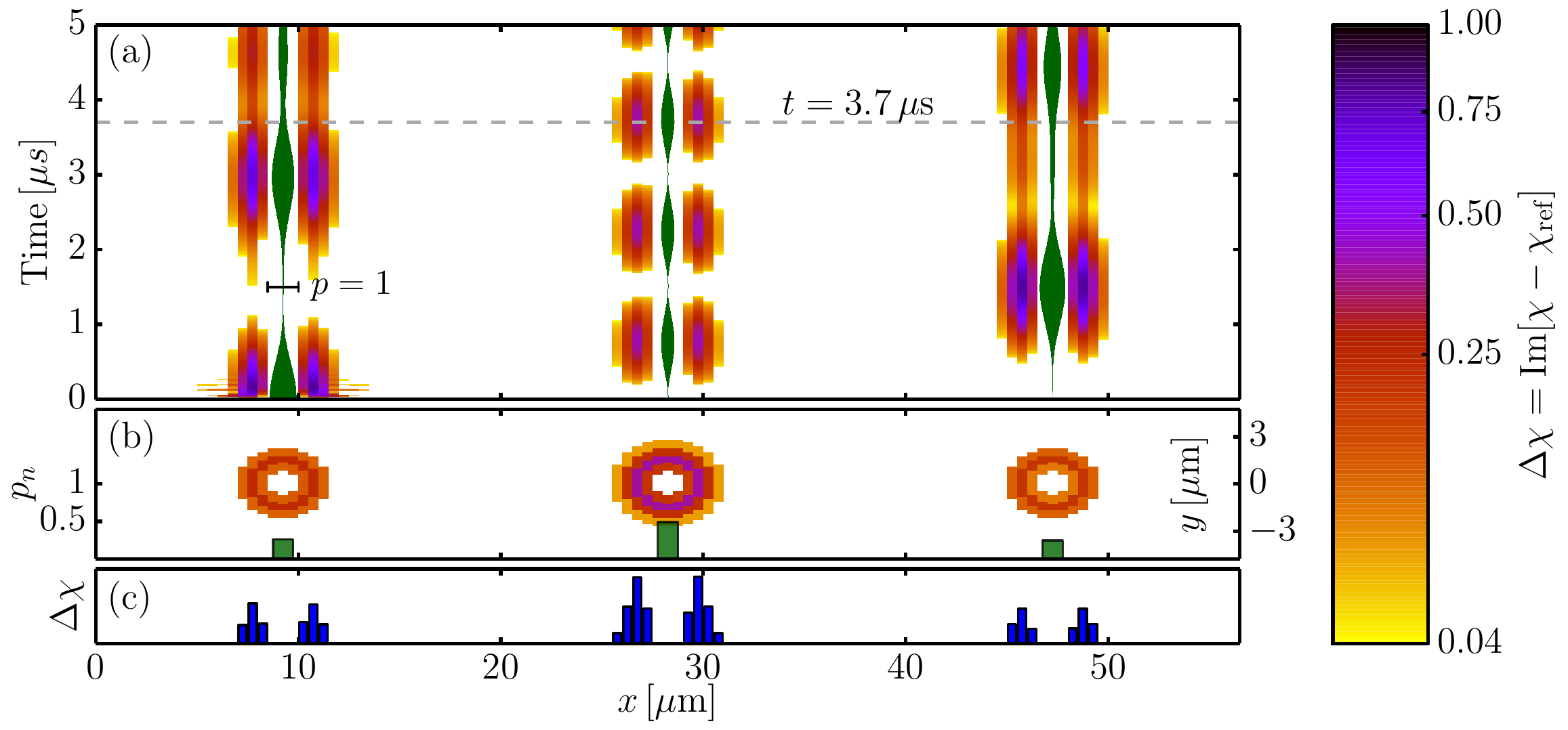,width= \figurewidth\columnwidth} 
\caption{(color online) Link between absorption signature and excitation transport. (a) Difference in optical response between dynamic and empty aggregate 
${\Delta}\chi(x)= \mathrm{Im}[\chi-\sub{\chi}{ref}]$ during transport, normalized by the two-level ($\ket{g}$, $\ket{e}$) response. Green lines indicate 
the location of aggregate atoms $r_n$ and their thickness the population $p_n$. (c) Snapshot of ${\Delta}\chi(x)$ at $t=3.7{\mu}s$. (b) Visualization of the 
corresponding two-dimensional signal \cite{footnote:2dextrap}, green bars indicate $p_n$. Here, $\Omega_p=0.2$ MHz, otherwise parameters as in 
\fref{many_body_simulations}. 
\label{chi_ofx_frames}}
\end{figure}
Each background atom $\alpha$ heralds the arrival of an aggregate excitation through the optical susceptibility $\chi_\alpha(t)=\Gamma_p/\Omega_p$Tr$(\hat{\rho} 
\hat{\sigma}^{(\alpha)}_{eg})$, the imaginary part of which yields the optical absorption. 
The average optical susceptibility of the background gas $\chi(x)$ is approximated by spatial binning of the $\chi_\alpha(t)$ from many simulations. To monitor 
the excitation transport, one can infer the location of the $\ket{p}$ state by subtracting from $\chi(x)$ a reference signal $\sub{\chi}{ref}(x)$ corresponding 
to the absorption of an inactive aggregate (chain of only $\ket{s}$ states) as in \cite{guenter:EITexpt}. 
We see in \fref{chi_ofx_frames} that the resulting signal is directly linked to the probability distribution of the excitation $p_{n}(t) = $Tr$(\hat{\rho} [\ket{\pi_n} \bra{\pi_n}])$. These simulations also show that 
background-background interactions $V_{\alpha\beta}^{(rr)}$ are relatively benign for the chosen states and densities.

\begin{figure}[t!]
\centering
\epsfig{file=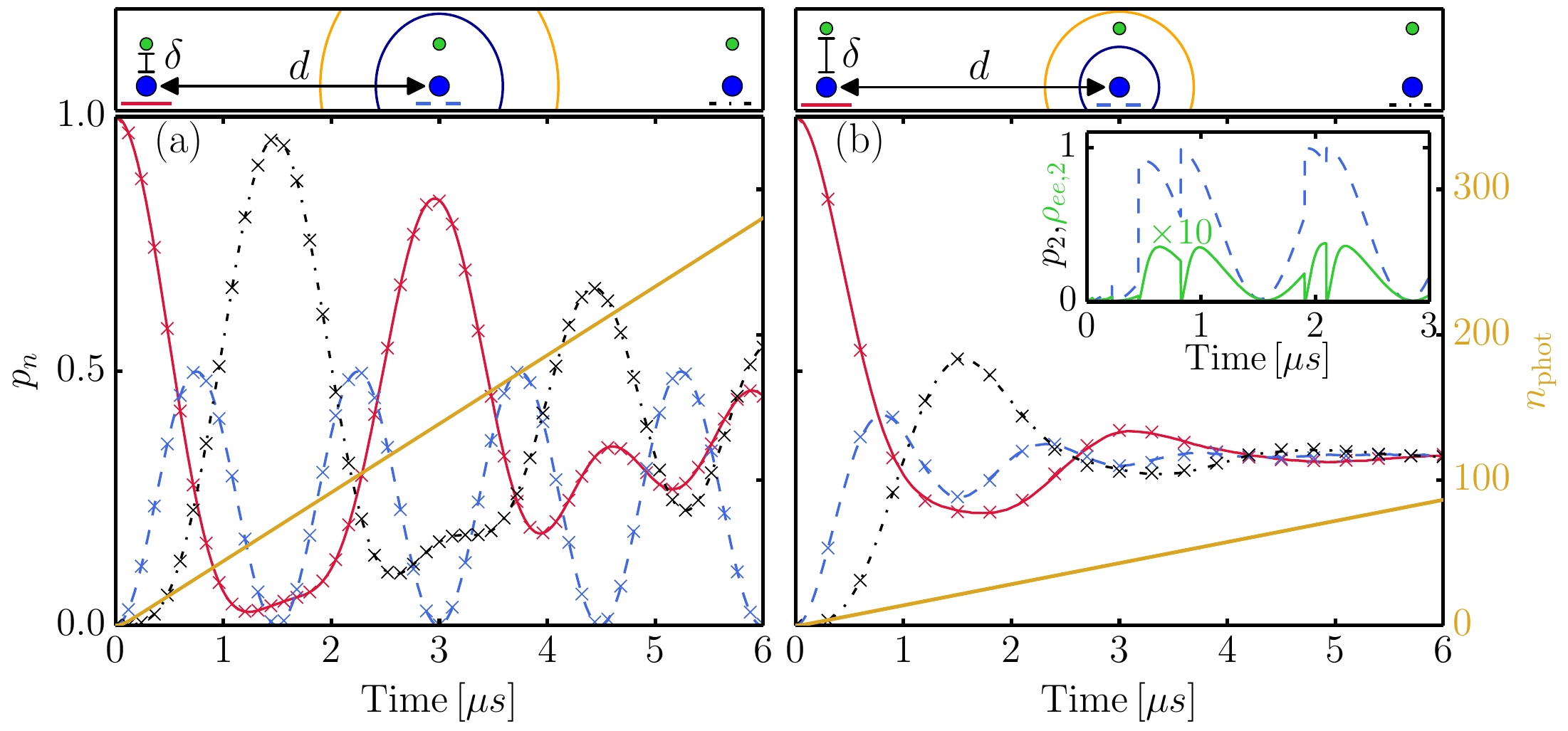,width= \figurewidth\columnwidth} 
\caption{(color online) Exciton transport in a continuously monitored embedded Rydberg aggregate. Geometries are shown in the top panels.  (a) Site occupations 
$p_n$ for $n=1,2,3$ (solid red, dashed blue, dot-dashed black) in a non-decohering case, $\Omega_p=1.3$ MHz, $\Omega_c=30$ MHz, $d=19 \mu$m, 
$\delta=0.6 \mu$m, other parameters as in \frefp{long_chain}{a}. Colormatched crosses show the populations according to the effective model, 
\eref{effective_eqn}. The orange line counts the number of scattered photons  $\sub{n}{phot}(t)=\int dt  \Gamma_p \sum_\alpha $Tr$(\hat{\rho}(t) [\ket{e} 
\bra{e}]_\alpha)$. (b) The same for a strongly decohering case with $\delta=1.5 \mu$m, other parameters as in (a).
\label{many_body_simulations}}
\end{figure}

The dephasing of the aggregate depends strongly on the position of the background atoms. In particular a given background atom only provides 
significant information on the excitation location if it is located in a ring \emph{between} the two critical radii $R_{c,s}<r<R_{c,p}$ as 
visible in \fref{chi_ofx_frames}. This is demonstrated in \fref{many_body_simulations}  where we place one background atom at a distance $\delta$ from each site as 
shown in the top panels. For $\delta<R_{c,s}$ background atoms permanently scatter a large number of photons, nonetheless the aggregate dynamics proceeds 
coherently (panel a). In contrast, for $R_{c,s}<\delta<R_{c,p}$, despite a smaller total number of scattered photons, aggregate decoherence is strong. 

The connection between information provided by 
the scattered photon and decoherence is explicit in the quantum-jump algorithm: The inset of \frefp{many_body_simulations}{b} shows for a single realization 
how the state of the aggregate $p_2$ (blue dashed) is linked to quantum jumps of the $\ket{e}$ population of its probe atom (green). This link only occurs when 
the state of the background atom and the state of the aggregate are significantly entangled in the moment of spontaneous decay. Since this is not the case in panel 
(a), single trajectories (not plotted) there show no effect of quantum jumps on the state of the aggregate. 

\ssection{Conclusions and outlook}
%
We have shown that a Rydberg aggregate embedded in an optically coupled background gas realizes a flexible quantum simulator of a Haken-Reineker-Strobl type model 
for energy transport. Site-dependent dephasing and disorder can be controlled through laser intensities, frequencies and background atomic density. 
Furthermore, this system could be extended to study other fundamental features believed to be at play in photosynthetic 
light-harvesting:  We have seen evidence for non-Markovian features and non-trivial relaxation when the time scale on which the 
background atoms reach their steady state is made comparable to transport time scales, a regime not discussed here.
The analogue of internal molecular vibrations could be engineered as in \rref{hague:Su_Schrieffer_Heeger}. Disorder distributions could be controlled even 
further using an additional class of  background atoms \cite{moebius:levydisorder}. All these features would extend the HRS type model 
proposed here to quantum simulations of light-harvesting processes in a similar spirit but with complementary technology to the proposals of \rref{herrera:holstein_pra,herrera:holstein,mostame:supercond}. 

Decoherence of the aggregate arises through continuous monitoring of the location of the excitation, providing a hands-on example of measurement-induced 
decoherence of a quantum state. Further applications of this system could be monitoring and decoherence of adiabatic excitation transport involving external 
(motional) degrees of freedom \cite{wuester:cradle,moebius:cradle}.

\acknowledgments

We gladly acknowledge fruitful discussions with Cenap Ates, Rick van Bijnen, Martin G{\"a}rttner, Georg G{\"u}nter, Igor Lesanovsky, Sebastian M{\"o}bius, Thomas Pohl, Hanna Schempp and Matthias Weidem{\"u}ller. We acknowledge financial support by the Deutsche Forschungsgemeinschaft under
WH141/1-1 and the EU Marie Curie Initial Training Network (ITN) COHERENCE.

\section{Supplemental information}

\ssection{Steady state of an EIT system}
For a single aggregate atom in state $a$ interacting with one background atom \cite{guenter:EIT} at a distance $\delta$, consider the Hamiltonian of that 
background atom,
\begin{align}
\sub{\hat{H}}{EIT}  &= \left(\frac{\sub{\Omega}{p}}{2}\hat{\sigma}^{(\alpha)}_{eg}  + \frac{\sub{\Omega}{c}}{2}\hat{\sigma}^{(\alpha)}_{re}  + 
\mbox{H.c.}\right) - \Delta \hat{\sigma}^{(\alpha)}_{rr}. 
\label{Hdet}
\end{align}
The detuning $\Delta$ will in this case be given by the interaction with the aggregate atom as $\Delta = V^{(ra)}(\delta)$.

We can solve the corresponding master equation including spontaneous decay from state $\ket{e}$ for its steady state $\tilde{\rho}$ and obtain
\begin{align}
\tilde{\rho}_{gg}(\Delta)&=\frac{4 \Gamma_p^2 \Delta^2 + \Omega_c^4 + (4\Delta^2 + \Omega_c^2) \Omega_p^2}{4 \Gamma_p^2 \Delta^2 + 8 \Delta^2 \Omega_p^2 + (\Omega_p^2 + \Omega_c^2)^2},
\\
\tilde{\rho}_{ee}(\Delta)&=\frac{4 \Delta^2 \Omega_p^2}{4 \Gamma_p^2 \Delta^2 + 8 \Delta^2 \Omega_p^2 + (\Omega_p^2 + \Omega_c^2)^2},
\label{SS_rhoee}\\
\tilde{\rho}_{rr}(\Delta)&=\frac{\Omega_p^2 (\Omega_p^2 + \Omega_c^2)}{4 \Gamma_p^2 \Delta^2 + 8 \Delta^2 \Omega_p^2 + (\Omega_p^2 + \Omega_c^2)^2},
\label{SS_rhorr}\\
\tilde{\rho}_{ge}(\Delta)&=\frac{2 \Delta \Omega_p (\Omega_c^2 -2 i \Gamma_p \Delta)}{4 \Gamma_p^2 \Delta^2 + 8 \Delta^2 \Omega_p^2 + (\Omega_p^2 + \Omega_c^2)^2},
\\
\tilde{\rho}_{gr}(\Delta)&=-\frac{\Omega_p \Omega_c (\Omega_p^2  + \Omega_c^2 -2 i \Gamma_p \Delta)}{4 \Gamma_p^2 \Delta^2 + 8 \Delta^2 \Omega_p^2 + (\Omega_p^2 + \Omega_c^2)^2},
\\
\tilde{\rho}_{er}(\Delta)&=-\frac{2 \Delta \Omega_p^2  \Omega_c}{4 \Gamma_p^2 \Delta^2 + 8 \Delta^2 \Omega_p^2 + (\Omega_p^2 + \Omega_c^2)^2}.
\end{align}
We further define the steady state susceptibility
\begin{align}
\tilde{\chi}(\Delta)&= \frac{\Gamma_p}{\Omega_p} \mbox{Im}[ \tilde{\rho}_{ge}(\Delta)].
\end{align}
This expression can be used to describe the time-dependent absorption signal in Fig. 3: 
For $W(d)\ll \Gamma_p$, we find that each background atom $\alpha$ adiabatically follows the aggregate state through 
$\chi_\alpha(t)=\sub{\chi}{adiab,$\alpha$}(t)= \sum_n \tilde{\chi}(\bar{V}_{n\alpha})p_{n}(t)$, where  $\bar{V}_{n\alpha} = V^{(rp)}_{n\alpha}+ \sum_{m\neq n} V^{(rs)}_{m\alpha}$ is the overall interaction of the specific background atom $\alpha$ with the entire aggregate if the latter is in the state $\ket{\pi_n}$. 

\ssection{Adiabatic elimination of background atom excited states}
Following \cite{sorensen:adiabelim} we now formally adiabatically eliminate the excited states of all background atoms to arrive at an evolution equation for the aggregate alone.
The essential step is to divide the (many-body) Hilbert space into a space of interest and its complement. The former is represented by the projector
\begin{align}
\hat{P}_g &= \sum_n \ket{\pi_n}\bra{\pi_n}\otimes \ket{\bv{g}}\bra{\bv{g}},
\label{Pg}
\end{align}
where the first part acts on the state space of the aggregate atoms and the second on that of the background atoms. We introduced $\ket{\bv{g}}=\ket{g\dots g}$, 
which is the state where all background atoms are in $\ket{g}$.
The complement of the space projected onto by $\hat{P}_g$ is thus formed by all many-body states involving any $\ket{e}$ or $\ket{r}$ state for the background 
atoms, projected onto by $\hat{P}_e=1-\hat{P}_g$.

Segregating the total Hamiltonian from the main article into segments using the projection operator formalism \cite{sorensen:adiabelim} to first order in $\Omega_p$, we obtain: 
\begin{align}
\hat{H}_g &= \hat{P}_g\hat{H}\hat{P}_g = \hat{P}_g\sub{\hat{H}}{agg}\hat{P}_g,
\label{Hg}
\\
\hat{H}_e &= \hat{P}_e\hat{H}\hat{P}_e 
\CR &
=\hat{P}_e  \bigg[  \sum_\alpha  (\frac{\sub{\Omega}{c}}{2}\hat{\sigma}^{(\alpha)}_{re}  + \mbox{H.c.}  ) -\Delta_p \hat{\sigma}^{(\alpha)}_{ee} 
\CR &
-(\Delta_p+\Delta_c) \hat{\sigma}^{(\alpha)}_{rr}+  \sub{\hat{H}}{int} +  \sub{\hat{H}}{agg} \bigg]  \hat{P}_e,
\label{He}
\\
\hat{C}_+ &=   \hat{P}_e \bigg[  \sum_\alpha \frac{\sub{\Omega}{p} }{2}\hat{\sigma}^{(\alpha)}_{eg} \bigg]  \hat{P}_g,  \:\:\:\:\:\:\:\:\:\:\:\: \hat{C}_- =\hat{C}_{+}^{\dagger}.
\label{VV}
\end{align}

After this segregation, the effective equation after adiabatic elimination of the complement of our space of interest is \cite{sorensen:adiabelim}:
\begin{align}
\supp{\dot{\hat{\rho}}}{(agg)}&=-i[\sub{\hat{H}}{agg} + \sub{\hat{H}}{eff},\supp{\hat{\rho}}{(agg)}]+
 \sum_\alpha {\cal L}_{\sub{\hat{L}}{eff}^{(\alpha)}  } [\supp{\hat{\rho}}{(agg)}],
\label{effeqns}
\\
\sub{\hat{H}}{eff}&=-\frac{1}{2} \hat{C}_- [\sub{\hat{H}}{NH}^{-1} + (\sub{\hat{H}}{NH}^{-1})^\dagger  ] \hat{C}_+  + \hat{H}_g,
\label{Heff}
\\
\sub{\hat{L}}{eff}^{(\alpha)} &=\hat{L}_\alpha \sub{\hat{H}}{NH}^{-1} \hat{C}_+. 
\label{Leff}
\end{align}
Here $\hat{L}_\alpha$ is the Lindblad operator introduced in the main article. Central to the effective equation are the non-Hermitian Hamiltonian $\sub{\hat{H}}{NH}=\hat{H}_e - i \sum_\alpha \hat{L}^\dagger_\alpha  \hat{L}_\alpha/2$ and its inverse $\sub{\hat{H}}{NH}^{-1}$, which we obtain now. 

It can be seen that if we neglect $V_{\alpha\beta}^{(rr)}$ and $\sub{\hat{H}}{agg}$ as we will do from now, the Hamiltonian $\sub{\hat{H}}{NH}$ decomposes into 
the following block structure:
\begin{align}
\sub{\hat{H}}{NH}&=\bigotimes_{n\alpha} \left[ \ket{\pi_n}\bra{\pi_n}\otimes \sub{\hat{H}}{NH}^{(n\alpha)}\right], 
\label{Hnhblocks}
\end{align}
where $\sub{\hat{H}}{NH}^{(n\alpha)}$ acts within the space spanned by $\ket{e}_\alpha$ and $\ket{r}_\alpha$ only. Similar blocks arise in $\hat{C}_+$ and $\hat{L}_\alpha$.

In that basis, $\sub{\hat{H}}{NH}^{(n\alpha)}$ reads explicitly:
\begin{align}
\sub{\hat{H}}{NH}^{(n\alpha)}&=
\left(
\begin{array}{cc}
-i \Gamma_p/2 -\Delta_p & \Omega_c/2\\
 \Omega_c/2 & -\Delta_p-\Delta_c + \sub{\bar{V}}{$n\alpha$} 
\end{array}
\right),
\label{HNH_oneblock}
\end{align}
where  $\bar{V}_{n\alpha} = V^{(rp)}_{n\alpha}+ \sum_{m\neq n} V^{(rs)}_{m\alpha}$ is the overall interaction of the specific background atom $\alpha$ with the entire aggregate if the latter is in the state $\ket{\pi_n}$. We further have
\begin{align}
\hat{C}_+^{(n\alpha)}&=
\left(
\begin{array}{c}
\Omega_p\\
0
\end{array}
\right),\:\:\:
\hat{L}^{(n)}_\alpha=(\sqrt{\Gamma_p}, 0).
\label{VLblocks}
\end{align}

Due to the block structure \bref{Hnhblocks}, we find the inverse of $\sub{\hat{H}}{NH}$ when we find the inverse of $\sub{\hat{H}}{NH}^{(n\alpha)}$, which is:
\begin{align}
&(\sub{\hat{H}}{NH}^{(n\alpha)})^{-1}=
 \CR
 &
\left(
\begin{array}{cc}
\tilde{V}_{n\alpha} & - \Omega_c/2 \\
- \Omega_c/2 & - i \Gamma_p/2-\Delta_p
\end{array}
\right)/\left[(-i\frac{\Gamma_p}{2}  -\Delta_p)  \tilde{V}_{n\alpha} - \frac{\Omega_c^2}{4} \right],
\label{HNH_oneblock_inverse}
\end{align}
with $\tilde{V}_{n\alpha}=\bar{V}_{n\alpha}-\Delta_p -\Delta_c$

Using also \bref{VLblocks}, we obtain
\begin{align}
\supp{\dot{\hat{\rho}}}{(agg)}&=-i[\sub{\hat{H}}{agg} + \sub{\hat{H}}{eff},\supp{\hat{\rho}}{(agg)}]+
 \sum_\alpha {\cal L}_{\sub{\hat{L}}{eff}^{(\alpha)}  } [\supp{\hat{\rho}}{(agg)}],
\\
\sub{\hat{H}}{eff}&=   \sum_{n}  \sum_\alpha  \sub{H}{eff}^{(n\alpha)}\ket{\pi_n}\bra{\pi_n},
\CR
\sub{\hat{L}}{eff}^{(\alpha)}&= \sum_n  \sub{L}{eff}^{(n\alpha)} \ket{\pi_n}\bra{\pi_n},
\CR
\sub{H}{eff}^{(n\alpha)}&=
\frac{ \Omega_p^2  \tilde{V}_{n\alpha}   (\Omega_c^2+ 4  \tilde{V}_{n\alpha}  \Delta_p)}{\Omega_c^4 + 8  \tilde{V}_{n\alpha} \Delta_p\Omega_c^2 + 4  
\tilde{V}_{n\alpha} ^2 (\Gamma_p^2 + 4 \Delta_p^2)},
\label{Heffna}
\\
\sub{L}{eff}^{(n\alpha)}&=  \frac{2i  \tilde{V}_{n\alpha} \sqrt{\Gamma_p}\Omega_p}{2 \tilde{V}_{n\alpha} (\Gamma_p - 2i \Delta_p) -  i \Omega_c^2}
\label{Leffna}
\end{align}
in \eref{Heff} and \eref{Leff}. In the limit $\Delta_{p,c}\rightarrow0$, we arrive at the expressions (5) to (7) of the main article.

In the main article we point out that even strongly absorbing background atoms whose location however does not allow one to infer the excitation location do not contribute to aggregate decoherence.
This is fully captured in the model just derived, where Lindblad operators $\sub{\hat{L}}{eff}^{(\alpha)}$ in Eq.~(7) of the main article for background atoms that are 
within a critical radius regardless of the excitation location $n$ are proportional to a unit matrix and hence cause no decoherence (since $\hat{O}$ in the super-operator ${\cal L}_{\hat{O}}[\hat{\rho}]$ commutes with $\hat{\rho}$, as $\hat{O}\sim \mathbb{1}$).

In a more explicit form of the HRS-type master equation, defining $W_{nn}=0$, we have
\begin{align}
\dot{\rho}_{nm}&= \sum_{k} i (W_{km} \rho_{nk} - W_{nk} \rho_{km}) 
 \CR &
+ i (E_m-E_n+ \epsilon_{nm}) \rho_{nm} - \frac{ \gamma_{nm}}{2} \rho_{nm},
\CR
 \gamma_{nm}&= \sum_\alpha(|\sub{L}{eff}^{(n\alpha)}|^2 + |\sub{L}{eff}^{(m\alpha)}|^2 -2\mbox{\cal Re}[\sub{L}{eff}^{(n\alpha)}\sub{L}{eff}^{(m\alpha)*}] ),
\CR
E_n &= \sum_\alpha \sub{H}{eff}^{(n\alpha)},\:\:\:\:\:\: \epsilon_{nm}= \sum_\alpha\mbox{\cal Im}[\sub{L}{eff}^{(n\alpha)}\sub{L}{eff}^{(m\alpha)*}],
\label{explicitmasterequation}
\end{align}
for the matrix elements of $\supp{\hat{\rho}}{(agg)}=\sum_{nm}\rho_{nm}\ket{\pi_n}\bra{\pi_m}$. We can interpret $E_n$ as diagonal disorder, $\gamma_{nm}$  
as dephasing, and $\epsilon_{nm}$ as correction to the diagonal disorder, as explained in the next paragraph.

Note that \bref{effeqns} is obtained as leading order of a perturbative expansion in $\hat{C}_{\pm}$ and by assuming that $\hat{H}_g$ can be treated as 
``small'' compared to $\hat{H}_e$. Higher orders and corrections due to finite $\hat{H}_g$ within $\sub{\hat{H}}{NH}^{(n\alpha)}$ can be incorporated as 
described in \cite{sorensen:adiabelim} but have not been required here.

\ssection{Disorder correction $\epsilon_{nm}$}
Since it depends on two aggregate site indices $n,m$ in a non-trivial fashion, the interpretation of $\epsilon_{nm}$ is not obvious at first. However, numerical 
evaluation shows that for the parameters of Fig.~2~(a) in the main text, $\epsilon_{nm}$ can be approximated by
\begin{align}
 \epsilon_{nm} \sim 2\sum_\alpha \left.\left(\sub{H}{eff}^{(n\alpha)} - \sub{H}{eff}^{(m\alpha)} \right)\right|_{C_{4,rp}=0}\label{eps_H}.
\end{align}
The Gaussian distribution $P_\epsilon$ in Fig.~2~(a) is reproduced by this expression with a standard deviation overestimating the correct one by 
$\sim 10\%$. Thus, $\epsilon_{nm}$ can be approximated by sums of contributions arising from single aggregate sites, and
consequently the term $ i (E_m-E_n+ \epsilon_{nm}) \rho_{nm}$ in \eref{explicitmasterequation} can be cast into the form $i (E_m'-E_n') \rho_{nm}$, with $E_k' = E_k - 2\sum_\alpha 
\sub{H}{eff}^{(k\alpha)}|_{C_{4,rp}=0}$. This is the reason we refer to $\epsilon_{nm}$ as to a correction to the diagonal disorder.

The approximation \bref{eps_H} does not hold true for all sets of parameters, in particular not for those of Fig.~2~(b). However, there
$\epsilon_{nm}$ is negligible.

\ssection{Dephasing in the continuum limit}
The dephasing rates $\gamma_{nm}$ in \eref{explicitmasterequation} are obtained as a discrete sum over all background atoms $\alpha$.
In the limit of a continuous background atom density (from here on denoted by $\rho_{\rm bg}$), these rates can be given as a 
closed analytical expression, provided that the interatomic distance of the aggregate satisfies $d\gg R_{c,rs},\,R_{c,rp}$
and that the probe and control fields are applied resonantly ($\Delta_p=0=\Delta_c$). In this continuum limit $\gamma_{nm}=\gamma (1- \delta_{nm})$
actually becomes site independent.

The dephasing $\gamma$ acquires three contributions, one
depending solely on the $rs$-interaction, one depending solely on the $rp$-interaction, and one depending on both:
\begin{equation}
\gamma=\gamma_{rs}+\gamma_{rp}+\gamma_{rs,rp}.
\end{equation}
The first two contributions have the simple form
\begin{eqnarray}
\gamma_{rs}&=&\frac{4\pi^2\Omega_p^2}{3\sqrt{2}\Gamma_p}\rho_{\rm bg} R_{c,rs}^3,\\
\gamma_{rp}&=&\frac{\pi^2\Omega_p^2}{\cos(\pi/8)\Gamma_p}\rho_{\rm bg} R_{c,rp}^3,
\end{eqnarray}
while the third contribution reads
\begin{widetext}
\begin{eqnarray}
\nonumber
&&\gamma_{rs,rp}=\frac{4 \pi^2 C_{4,rp}^5 \Omega_p^2 \rho_{\rm bg}}{3 \sqrt[4]{\Gamma_p |C_{4,rp}^{21}|} (4 \Gamma_p^2 C_{4,rp}^6 
\Omega_c+C_{6,rs}^4\Omega_c^5)}\left\{2 \sqrt[12]{\Gamma_p C_{6,rs}^2 |C_{4,rp}|^{15}}\right.
\\
\nonumber
&\times&\left[(\sqrt{3}+1) \Gamma_p C_{4,rp}^3 \left(\sqrt[3]{16 \Omega_c^4
|C_{6,rs}|^5}+C_{4,rp} C_{6,rs} \sqrt[3]{2^5 \Gamma_p  \Omega_c^2}\right)\right.
- 2 C_{4,rp}^2 \left(2 C_{4,rp}^3 \sqrt[3]{\Gamma_p^5 |C_{6,rs}|}+C_{6,rs} \Omega_c^2 (\Gamma_p C_{6,rs}^2)^{2/3}\right)
\\
&+& \left. (1-\sqrt{3}) \left(C_{6,rs}^3 \sqrt[3]{2 \Omega_c^{10} |C_{6,rs}|^2} - C_{4,rp} |C_{6,rs}|^3 \sqrt[3]{4 \Gamma_p
\Omega_c^8}\right)\right]
\\
\nonumber
&-& \frac{3}{\sqrt[4]{2}} \left[2 \Gamma_p C_{4,rp}^3 C_{6,rs} \sqrt{\Omega_c} \csc\left(\frac{\pi}{8}\right) |C_{4,rp}|
\left(C_{4,rp} \sqrt{2 \Gamma_p |C_{4,rp}|}+ C_{6,rs} \Omega_c\right)\right.
\\
\nonumber
&+& \left. \left. C_{6,rs}^3\sqrt{\Omega_c^5} \sec\left(\frac{\pi }{8}\right) \left(C_{4,rp} C_{6,rs} \Omega_c-\sqrt{2 \Gamma_p  |C_{4,rp}|^5}\right)\right]\right\}.
\end{eqnarray}
\end{widetext}
A similar analytical solution can be found for the on-site energy shifts \bref{Heffna}, but since these also become independent of the site-index, and thus disorder vanishes, it is not shown here.



\begin{thebibliography}{79}
\bibitem[{foo({\natexlab{a}})}]{footnote:interactions}
\bibinfo{note}{For these quantum numbers we obtain dispersion coefficients
  $C_{6,rr}=3.67 \times 10^{18}$ au, $C_{6,rs}=-6 \times 10^{17}$ $C_{4,rp}=-2
  \times 10^{10}$ from exact diagonalisation of the di-atomic Hamiltonian, in
  agreement with \cite{singer:VdWcoefficients}. Due to the resonance $V_{rp}$
  has several significant dispersion coefficients, the $R^{-4}$ form has been
  chosen to reproduce the same critical radius $R_{c,p}$ as the full
  interaction for our parameters.}
\bibitem[{foo({\natexlab{b}})}]{footnote:cutoff}
\bibinfo{note}{For practical reasons we cut off state energies at
  $\sub{V}{max}=600$ MHz. This still amounts to a full Rydberg blockade.}
\bibitem[{foo({\natexlab{c}})}]{footnote:2dextrap}
\bibinfo{note}{The simulation only featured a one-dimensional distribution of
  background atoms. To visualize the expected experimental signal, we compute
  from the one-dimensional absorption signal around each aggregate atom
  $\sub{\chi}{1D}(\delta)=\sub{\chi}{1D}(x - x_n)$ the corresponding
  two-dimensional signal $\chi(x,y) = \chi(\bv{r}) \equiv
  \sub{\chi}{1D}(\delta)$, with $\delta = |\bv{r} - \bv{r}_n|$, where
  $\bv{r}=(x,y)^T$.}
\end{thebibliography}
\end{document}